# Modeling the Combined Impact of Rainfall and Storm Tide on Coastal Cities under a Changing Climate: Transportation Infrastructure Impacts in Norfolk, Virginia USA as a Case Study

Yawen Shen[a], Navid Tahvildari[b], Mohamed M Morsy[c], Chris Huxley[e], T. Donna Chen[f], Jonathan L. Goodall[g]

[a] Water Resources Engineer, 10205 Westheimer Rd #800, Houston, TX 77042. E-mail: ys5dv@virginia.edu

[b] Assistant Professor, Dept. of Civil and Environmental Engineering, Old Dominion University, Norfolk, VA 23529. E-mail: ntahvild@odu.edu

[c] Assistant Professor, Irrigation and Hydraulics Engineering Department, Faculty of Engineering, Cairo University., P.O. Box 12211, Giza 12614, Egypt. E-mail: mmm4dh@virginia.edu

[e] Senior Engineer, BTM WBM Pty Ltd, Level 8, 200 Creek Street, Brisbane, 4000, Australia. E-mail: chris.huxley@tuflow.com

[f] Assistant Professor, Dept. of Engineering Systems and Environment, Univ. of Virginia, Olsson Hall, Charlottesville, VA 22904. E-mail: tdchen@virginia.edu

[g] Professor, Dept. of Engineering Systems and Environment, Univ. of Virginia, Olsson Hall, Charlottesville, VA 22904 (corresponding author). E-mail: goodall@virginia.edu

**Abstract**

Low-lying coastal cities across the world are vulnerable to the combined impact of rainfall and storm tide. However, existing approaches lack the ability to model the combined effect of these flood mechanisms, especially under climate change and sea level rise (SLR). Thus, to increase flood resilience of coastal cities, modeling techniques to improve understanding and prediction of the combined effect of these flood hazards are critical. To address this need, this study presents a modeling system for assessing the combined flood risk to coastal cities under changing climate conditions that leverages ocean modeling with land surface modeling capable of resolving urban drainage infrastructure within the city. The modeling approach is demonstrated in quantifying the future impact on transportation infrastructure within Norfolk, Virginia USA. A series of combined storms events are modeled for current (2020) and projected future (2070) climate conditions. Results show that pluvial flooding causes a larger interruption to the transportation network compared to tidal flooding under current climate conditions. By 2070, however, tidal flooding will be the dominant flooding mechanism with even nuisance flooding expected to happen daily due to SLR. In 2070, nuisance flooding is expected to cause a 4.6% total link close time (TLC), which is more than two times that of a 50-year storm surge (1.8% TLC) in 2020. The coupled flood model was compared with a widely used but physically simplistic bathtub method to assess the difference resulting from the more complex modeling presented in this study. Results show that the bathtub method overestimated the flooded area near the shoreline by 9.5% and 3.1% for a 10-year storm surge event in 2020 and 2070, respectively, but underestimated flooded area in the inland region by 9.0% and 4.0% for the same events. The findings demonstrate the benefit of sophisticated modeling methods, beyond more simplistic bathtub approaches, in climate adaptive planning and policy in coastal communities.

# 1. Introduction

Coastal cities are increasingly vulnerable to flooding exposure owing to growing population, urbanization, climate change, and relative sea level rise (SLR) (Hanson et al., 2011; Hallegatte et al., 2013; Aerts et al., 2014, Neumann et al., 2015; Dawson et al., 2016). Flooding can be destructive in these low-lying, densely populated, and highly developed regions (Gallien et al., 2014; Karamouz et al., 2017). For example, Hurricanes Katrina (2005), Sandy (2012), Harvey (2017), and Florence (2019) caused significant loss of life and damage in the U.S. Gulf and Atlantic Coasts. According to Hanson et al. (2011), about 40 million people were exposed to a 100-year coastal flood event in the 136 largest port cities in the world in 2005. By the 2070s, the population exposed to such an event could increase threefold. Hallegatte et al. (2013) estimated flood losses in a single year by 2050 to be more than 10 times those in 2005. Despite facing these hazards, protection measures in coastal cities have often been inadequate (Nicholls et al., 2008), primarily due to short-term economic decisions and uncertainty of future risk (Aerts et al., 2014). Therefore, to help inform policy decisions, it is necessary to enhance the understanding and modeling capacities for assessing flood hazards in coastal cities under a changing climate.

Individual flood mechanisms, like pluvial flooding and tidal flooding, can cause widespread impacts in coastal cities (Comer et al., 2017; Mignota et al., 2019). However, if multiple mechanisms occur concurrently, flood severity can be greatly exacerbated (Xu et al., 2014; Wahl et al., 2015; Batten et al., 2017; Shen et al., 2019). Prior studies have demonstrated the statistical dependence of rainfall and storm tide in coastal regions (Wahl et al., 2015; Batten et al., 2017; Xu et al., 2019). Batten et al. (2017) analyzed the temporal dependence of rainfall and storm tide in Virginia Beach, Virginia and found that the correlation coefficients between rainfall and storm tide vary between 0.336 and 0.452 across different data sources. Meanwhile, they indicated that over half of the rainfall events occurred while the tide level was higher than the average high tide. Wahl et al. (2015) estimated the likelihood of joint occurrence of rainfall and storm tide across the contiguous United States (US) coast and concluded that the possibility of a compound storm is higher on the Atlantic/Gulf coast than on the Pacific coast. Meanwhile, they found that the frequency of compound storms has significantly increased in the past century, and this trend is expected to continue due to climate change. Xu et al. (2019) estimated the bivariate return periods of compound rainfall and storm tide based on copula functions and failure probability and showed significant correlation between rainfall and storm tide.

Given the correlation between storm tide and rainfall, considerable efforts have been made to develop methodologies for modeling the combined impact from these flood mechanisms. In prior work, the proposed modeling systems are normally the coupling of a one-dimensional (1D: Ray et al., 2011; Bacopoulos at al., 2017; Karamouz at al., 2017) or two-dimensional (2D: Yin et al., 2016; Silva-Araya et al., 2018) overland model with a 2D or three-dimensional (3D) storm surge model (e.g., ADCRIC, MIKE21, and Delft3D). Yin (2016) proposed a coastal inundation model by coupling a storm surge model (ADCIRC) with an urban flood model (FloodMap). At the city scale, the coupled model demonstrates improved results over ADCIRC modeling alone for both flood extent and depth. In Silva-Araya et al. (2018), a 2D hydrologic model was coupled with a 2D storm surge model for Puerto Rico. In an execution, the storm surge model will run first to prepared tidal boundary for the hydrologic model. Results show that the interaction between pluvial flooding and tidal flooding caused increased flooding compared to storm surge alone. Coastal regions are often located in low-relief terrain without a large amount of storage potential. Furthermore, the topography complexity increases significantly in urban environments due to the surface and subsurface infrastructure. Routing water in such environments requires high-resolution 2D hydrodynamic to model the complex street-level overland flooding (Karamouz at al., 2017; Shen at al., 2019). In urban environments, subsurface drainage systems play a key role in managing urban flooding; however, in prior work, they are often not explicitly included in the modeling systems (Yin et al., 2016; Silva-Araya et al., 2018). This is primarily due to the difficulty in obtaining a large-scale drainage system database in order to parameterize the model. Without explicitly including subsurface drainage systems, flood models may inaccurately estimate the effect of not only pluvial flooding, but also tidal flooding due to the backing up of ocean water into the city through storm drainage pipes during storm surge events. Therefore, to increase the accuracy of coastal city flooding, this study used an overland flood model consisting of a large-scale 1D subsurface drainage network and a 2D surface overland hydrodynamic model. This model system is capable of simulating the dynamics of surface runoff and pipe flow, as well as the interaction between them. The model was then coupled with a physics-based storm surge model built in Tahvildari and Castrucci (2020) in order to obtain ocean boundary conditions.

Methodologies with varying complexity have been developed in prior works to estimate the flood impact to critical infrastructure systems such as transportation infrastructure under climate change and SLR (Suarez et al., 2005; Arnbjerg-Nielsen et al., 2013; Sadler et al., 2017; Yin et al., 2016 and 2017). One simplified approach, the so called "bathtub" method, assumes that the water surface is level so that flood inundation can be easily estimated by using a digital elevation model (DEM). Because of its algorithmic simplicity and computational efficiency, the bathtub method is commonly applied to evaluate the impact of SLR and/or storm surge in coastal environments. Sadler et al. (2017), for example, used the bathtub method and traffic information to identify critical roadways in Norfolk and Virginia Beach, Virginia exposed to future flood hazards. A similar method was adopted by Jacobs et al. (2018) to assess the increasing vulnerability of roadways to nuisance flooding on the U.S. east coast. White the bathtub method is convenient, past studies suggest that it tends to overestimate flood depths and extents because of its physical oversimplification of flood routing (Ramirez et al., 2016; Castrucci and Tahvildari, 2018).

The objective of this study, therefore, is to advance methods for modeling the combined impact of rainfall and storm tide in coastal cities under a changing climate. As a demonstration, the method was applied for a portion of Norfolk, Virginia to assess the flood impact on the transportation network for a series of compound storm scenarios under the current and future climate conditions. The results are compared to the widely used bathtub approach to quantify the advantage of complex dynamic flood models in coastal urban environments. The main contribution of this study, therefore, is to advance the method for dynamic modeling the combined impact of rainfall and storm tide in coastal cities and demonstrate the application of this method in assessing flooding risk under both current and future projected climate conditions. The research also contributes (1) an investigation of the flood hazard to the transportation network in the case study region; (2) a quantification of improvements over more simplified bathtub modeling approaches, and (3) new approaches to evaluate high-resolution, city-scale flood model using drone footages.

## 2. Methodology

*2.1 Study Area*

The proposed methods are demonstrated in a portion of the City of Norfolk, Virginia, USA. Norfolk is in the heart of the Hampton Roads metropolitan area in southeast Virginia, and it is the home of the world's largest naval base. This highly urbanized and low-lying city is located at the confluence of the James River, Elizabeth River, and Chesapeake Bay. Norfolk faces a series of natural hazard challenges, including flooding and other impacts of climate change, SLR, and subsidence (Boon, 2012; Sweet et al., 2017). This region is experiencing the highest rate of relative sea level rise (RSLR) on the US Atlantic Coast (Boon et al., 2012; Koop, 2013). Studies have shown that flood frequency in this region has significantly increased due to SLR in the past decades, and this trend is expected to continue in coming decades (Moftakhari et al., 2015; Sweet et al., 2015; Burgos et al., 2018). Burgos et al. (2018) indicates that nuisance flooding has increased 325% in Norfolk since 1960. More frequent extreme storm surge events have been observed in Norfolk over the past two decades that 17 extreme storm surge events have occurred in the past 85 years in Norfolk, but more than half of these events occurred in the past 15 years (Ezer and Atkinson, 2014). Meanwhile, because of climate change, the intensity of a 24-hour duration rainfall with a greater than 1-year return period is expected to increase in Norfolk (VB., 2018; Morsy et al., 2019). Studies have shown that rainfall intensity in Norfolk is projected to increase 30% to 40% by 2075 compared to 2000 levels under the high emission scenario (VB., 2018; Morsy et al., 2019).

*2.2 Storm Surge Model*

The coupled hydrodynamic and wave model developed by Tahvildari and Castrucci (2020) was adopted to simulate storm tide under the current and future projected sea level conditions. The model is built on the Delft3D modeling suite (https://oss.deltares.nl/web/delft3d). Because storm surge is a shallow water process and the flow can be assumed well-mixed in the vertical direction, the model is applied in a 2D or depth-averaged mode with a single vertical layer.

Considering the computational cost of the storm surge simulation, models are normally designed to have a relatively coarser resolution at open sea and a higher resolution near shorelines, especially near critical regions. However, the computational efficiency is likely low when using a single grid crossing a large geographic region. Because the simulation time step is determined by the smallest cells in the grid, one approach to reduce the computational time is to set up the storm surge model in a nested structure. Tahvildari and Castrucci (2020) nested the storm surge model at two levels, where low-resolution Model 1 covered most of the Chesapeake Bay, the Eastern Shore, and areas immediately offshore the Chesapeake Bay in the Atlantic Ocean, and the high-resolution Model 2 covered most of the urban areas at the south of the Chesapeake Bay. Model 1 used an equidistant grid with cell size 125m×200m while Model 2 used a curvilinear grid with resolution of 10 m near critical infrastructures and 30-90 m away from these spots. Figure 1 shows the domains of the storm surge model and its relative location to the urban flood model introduced in the following section. During simulation, the velocity and water level estimates from Model 1 are transferred to Model 2 as boundary conditions.

The publicly available topography and bathymetry data from the National Oceanographic and Atmospheric Administration (NOAA) were used to build the model. The bathymetry and topography used in Models 1 and 2 were obtained from the Coastal Relief Model (CRM; 90m horizontal resolution) and a regional dataset (10m to 30m horizontal resolution). The open boundary condition of Model 1 uses TPXO8 global tide model (Egbert and Erofeeva, 2002).

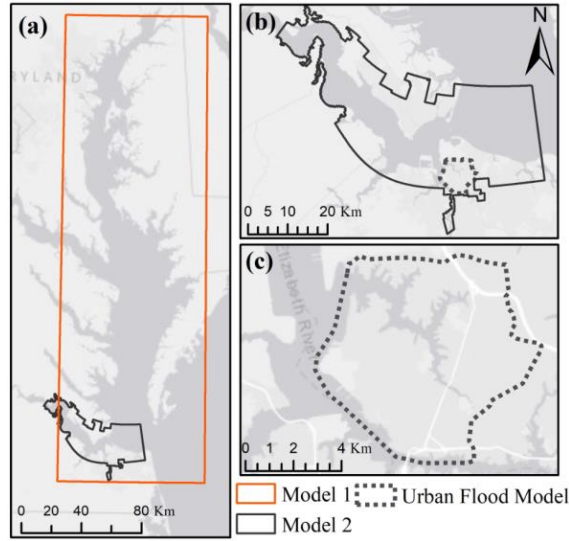

Figure 1. Storm surge model and urban flood model domains: (a) Model 1 (b) Model 2 (c) urban flood model

## 2.3 Urban Flood Model

The urban flood model that includes a 1D subsurface drainage network and a 2D surface domain was built for the southwest portion of Norfolk, as showed in Figure 2. The model domain has a total area of $56.4 km^2$, of which $8.7 km^2$ is open water. The study domain is a highly urbanized area with many artificial structures, including commercial and residential buildings, transportation networks, and stormwater infrastructure. The low-lying topography along with these structures create complex flow patterns and paths on both the land surface and in underground pipe networks. To accurately simulate street-level urban flooding, a 1D pipe/2D overland hydrodynamic flood model was built using the Two-dimensional Unsteady Flow (TUFLOW) model (Syme, 2001). TUFLOW was adopted because it is capable of simulating both surface flow on a 2D domain and pipe flow via its 1D functionality along with a dynamic link between the two domains. TUFLOW is also capable of leveraging Graphical Processing Units (GPUs) in parallel, which can significantly speed up the model simulations. In Shen et al. (2019), an urban flood model was built for the Hague Community located near downtown Norfolk. The spatial datasets, pipeline network, and parameter settings from Shen et al. (2019) were adopted to build the urban flood model for the larger study domain in this study.

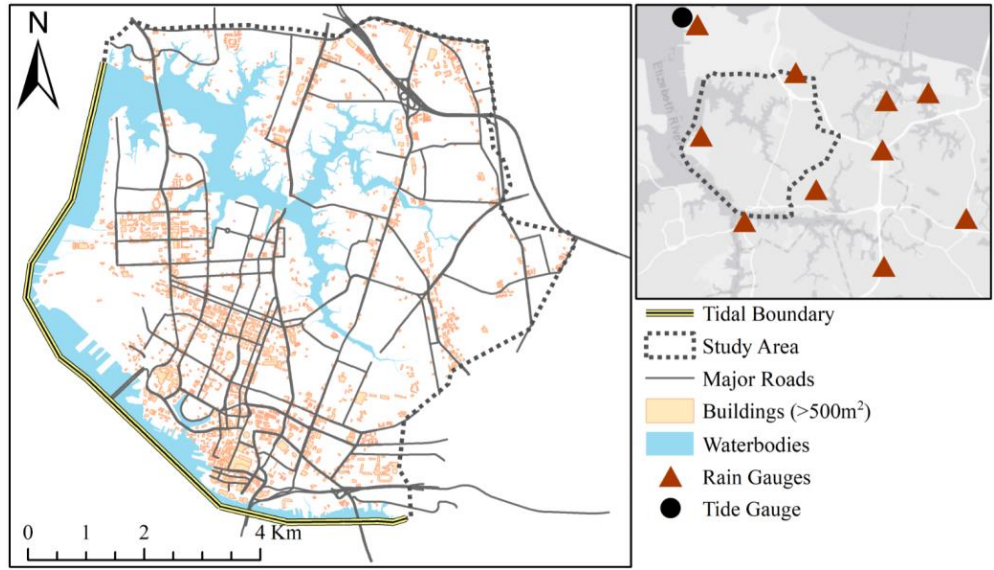

Figure 2. Urban flood model and locations of tide and rain gauges.

In this study, tide level observations were collected from the Sewells Point station (Station ID: 8638610) located along the northwest coast of Norfolk. The Sewells Points station, established in 1927, represents the longest tide record in Virginia. Rainfall observations were collected from two rain gauges maintained by the U.S. National Weather Service (NWS) and two rain gauges maintained by the Hampton Roads Sanitation District (HRSD). The locations of these gauges are shown in Figure 2.

The accuracy of topography representation is one of the most sensitive factors in the urban flood model. In this study, digital elevation model (DEM) representing the topography was built from a LiDAR-derived DEM dataset provided through the Virginia Geographic Information Network (VGIN) with 0.76m×0.76m horizontal resolution (Dewberry, 2014). This dataset, acquired in 2013, has a vertical accuracy of ±0.066m. Buildings are another major factor describing the landscape in urban environments. Building footprint and height information were also gathered from VGIN. In the study domain, there are more than 29,000 commercial and residential buildings. To reduce the complexity and increase numerical stability of the model, only buildings with areas greater than $500m^2$ were represented in the model. It was assumed that water will not enter the buildings, thus 3D building structures were added to the bare surface DEM.

The study domain has a complex drainage system. There are 9,087 pipe sections with a total length of about 222km. In this drainage system, more than half of the pipes were installed before 1950, and about 2,000 pipes were installed before 1920. Given the age and complexity of the system, it is challenging for the city to survey the entire pipeline system, and missing and inconsistent information may exists in the available surveyed data. We suspect this is a common challenged for other older cities as well. To reduce the complexity of the model, pipes with diameters of smaller than 15 inches were excluded. Pipes inside buildings, underpasses, and tunnels were also excluded. Major pipes with missing information were retained, and it was assumed that their missing properties were consistent with their upstream and downstream pipes. For example, a pipe with a missing invert elevation is assumed to have the same bed slope as its upstream and downstream pipes. Then, the upstream and downstream invert elevations of the pipe are calculated by interpolating the inverse distance weighting (IDW). Smaller pipes with missing information were excluded from the model. After these data cleaning steps, 6213 pipe sections were retained with the total length of about 179km.

TUFLOW solves the full 2D depth averaged momentum and continuity equations for shallow water free surface flow and incorporates the full functionality of the ESTRY one-dimensional (1D) hydrodynamic network model (Syme, 2001). The same technique and parameter settings for coupling the 2D land surface domain and the 1D pipe network used by Shen et al. (2019) were adopted by the current study. Please reference Shen et al. (2019) for details. The study domain has an average imperviousness ratio of 41% and a shallow groundwater level, which can reach the surface during storm events that cause flooding. Thus, infiltration is expected to not have a significant influence on significant flood events, and therefore, it was not parameterized in the urban flood model. This presumes saturated conditions for the non-impervious surfaces in the urbanized watershed at the start of the model simulation period. Lastly, detailed simulation of bridges, underpasses, and tunnels were excluded in the model to focus on other regions within the study domain. Future work can add additional complexity to the model to include these additional processes and features of the landscape.

*2.4 Model Coupling*

The storm surge model and the urban flood model were linked through the tidal boundary portion of the study domain, as shown in Figure 2. The total length of the tidal boundary is 12.5km. The tidal boundary was split into nine sections, which are approximately parallel to the regional coastlines with no sharp change in orientation. The length of these sections varies from 950m to 2000m. In a simulation, the storm surge model is executed first, separately from the urban flood model. Then, time series of tide level from the simulation are extracted and averaged across each section in the flood model using an automatic post-processing procedure. This step prepares the boundary condition of water stage versus time for each tidal boundary section. In the urban flood model, the initial tide level was set as the averaged low tide level in the past four decades, which is about -0.9m (NAVD88) at Sewells Point station. Then, the tide level boundary conditions and rainfall inputs are fed into the urban flood model to generate high-resolution flood simulations.

The flow velocity at the tidal boundary of the urban flood model is not included in the current version of the modeling system due to limitations in specifying velocity boundary conditions in the version of TUFLOW used in this study. This simplification would likely cause underestimation of storm tide inundation, but we do not suspect this underestimation to be significant. Nonetheless, in an effort to compensate for potential underestimation, a post-processing procedure was applied to generate the maximum inundation maps and the inundation maps for each time step. First, simulations from the storm surge model were extracted inside the urban flood model domain. Second, these storm surge simulations were resampled to the mesh of the urban flood model. Third, these resampled storm surge simulations were compared with the urban flood model simulations in each cell, and the maximum values were selected as the final result. This procedure was then repeated for each simulation time step. The procedure results in the final results that can be thought of as the worst-case situation from the two models, which are valuable for flood hazard mitigation purposes.

*2.5 Model Evaluation*

Direct flood observations in urban areas are rare; however, it is essential for urban flood model evaluation and calibration (Smith et al., 2012). A USGS gauge (latitude: 36.8588º, Longitude: -76.2986º) was installed in the Hague Community near the downtown Norfolk during Hurricane Irene (2011). Records from this gauge are used to assess the model performance at that location near the shoreline.

Spatially distributed flood observations, for example flood extent, have unique values for evaluating the performance of 2D models. Crowdsourced data, such as photos or camera footage of flooded areas, newspaper reports, personal interviews, and flooding information shared by users on social media, can be converted to inundation information (Smith et al., 2012; Middleton et al., 2014; Fohringer et al., 2015; Loftis et al., 2017). The inundation information is valuable for evaluating 2D urban flood models. In this study, imagery data was adopted as an additional innovative data source to evaluate the coupled flood model. On October 10th 2016, the day after the peak of Hurricane Matthew in Norfolk, drone footage was captured by the Norfolk Department of Emergency Management and is available online (https://www.youtube.com/watch?v=R8ZYxubUo-w). In the drone footage, three scenes were captured inside the current study domain. Because the flight information, such as flight height, orientation, camera angle, and precise geolocation, is missing from the record, the footage cannot be directly georeferenced. Therefore, we created a method to extract flood extent by manual flood edge identification and GIS processing.

As a demonstration, Figure 3 shows the procedure of the proposed method for extracting flood extent. The first step is to identify flood edges from multiple frames captured in the same scene. For example, in Figure 3a, the flood edge intersects with several features in the surrounding area, such as streetlamps, street curbs, and parking space lines. Flood edge points close to these features are manually selected and geolocated (Figure 3b). The following step is to generate elevation contour lines crossing each flood edge points by using the high-resolution LiDAR dataset (Figure 3c). The next step is to separately calculate the total distance of each contour line to all flood edge points. Among all the contour lines, the one with the minimum total distance is selected as the inundation edge (Figure 3d). Finally, the areas lower than the contour line elevation are assumed to be flooded at that time. Using the same procedure, the flood extents in all three scenes were extracted. One assumption of the procedure is that the water surface is level or nearly level, which is plausible for this case because the footages were captured one day after the storm peak.

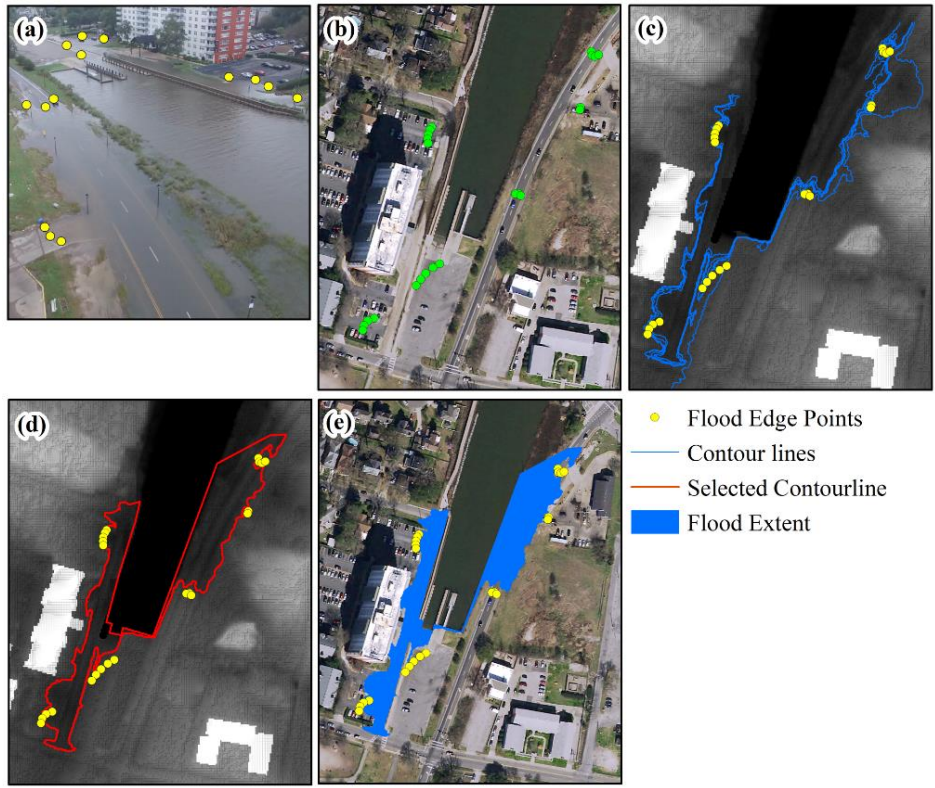

Figure 3. Procedure to extract flood extent from drone footages: (a) identify clear flood edge points; (b) manually geo-locate flood edge points on a world coordinate system; (c) generate elevation contour lines crossing each flood edge point; (d) select the contour line with the minimum total distance to all flood edge points; (e) determine the flood extent.

## 2.6 Compound Storm Scenarios

### 2.6.1 Relative Sea Level Rise Scenarios

Virginia's coast is a hot spot of RSLR due to climatic and non-climatic (e.g., subsidence) processes. The primary non-climatic component is vertical land movement (VLM; subsidence or uplift), but it also includes sea surface heigh changes associated with glacial isostatic adjustment (Sweet et al., 2017). In this study, the RSLR scenarios were obtained from Sweet et al. (2017). RSLR is equal to the sum of global mean sea level (GMSL) change and local relative sea level (RSL) change as follows

$$RSLR_{x,t} = \Delta GMSL_t + \Delta RSL_{x,t} \qquad (1)$$

where $RSLR_{x,t}$ is the total RSLR estimated for spatial location *x* and time *t*, $\Delta GMSL_t$ is the GMSL rise at time *t*, and $\Delta RSL_{x,t}$ is the total RSL change relative to RSL at the reference time ($t_0$). From Sweet et al. (2017), the RSL change that occurs in each GMSL rise scenario consists of contributions from climate-related processes and non-climatic background RSL changes. The total RSL change is defined as

$$\Delta RSL_{x,t} = Climatic\ \Delta RSL_{x,t} + Background\ RSL\ Rate_x(t - t_0) \qquad (2)$$

where Climatic $\Delta RSL_{x,t}$ is the RSL change affected by climate-related processes at spatial location *x* and time *t*, and $Background\ RSL\ Rate_x$ is the non-climatic component of RSL change, which is assumed to be linear in time.

We adopted the RSLR scenarios from Sweet et al. (2017), which revised the upper and lower bounds of SLR scenarios from the most up-to-date scientific literature. Sweet et al. (2017) recommends six GMSL rise scenarios from 0.3 m to 2.5 m by the year 2100. These six GMSL rise scenarios are Low, Intermediate-Low, Intermediate, Intermediate-High, High, and Extreme, which correspond to GMSL rise of 0.3 m, 0.5 m, 1.0 m, 1.5 m, 2.0 m, and 2.5 m, respectively. Based on Sweet et al. (2017), Tahvildari and Castrucci (2020) calculated the RSLR scenarios at the Sewells Point tide gauge with the mean background RSL rate of 2.47mm/yr. In this study, we adopted these RSLR projections, as shown in Figure 4. Note that these RSLR projections are referenced to zero at 2000.

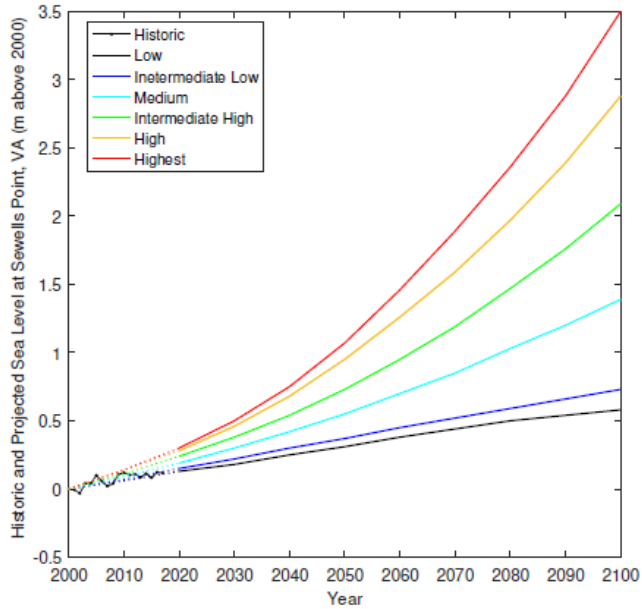

Figure 4. Historical and projected relative sea level at the Sewells Point tide gauge in Norfolk, VA (Tahvilidari and Castrucci, 2020)

In this study, 10 and 50-year storm surge events, based on water level observation at the Sewells Point tide gauge, are chosen to represent a moderate and a high storm surge, respectively. The storm surge elevations for the 10-year (1.55m NAVD88) and 50-year (2.07m NAVD88) recurrence intervals were obtained from the FEMA Flood Insurance Study for Norfolk (FEMA, 2017). The peak of Hurricane Irene (1.81 m, 2011) falls between the 10-year and 50-year tide levels under the current climate condition. The track and central pressure information of Hurricane Irene (2011), obtained from the NOAA National Hurricane Center (Avila and Cangialosi, 2011), were used to generate the pressure and wind field over the storm surge domain. Then, the wind intensity was adjusted to produce the 10 and 50-year water levels. For further details, please refer to Tahvilidari and Castrucci (2020). We selected the high RSLR scenarios as a demonstration for a conservative scenario. The high GMSL rise projection corresponds to a 2.0 m rise by 2100 with the probability of exceedance at 0.1% (RCP2.6) and 0.3% (RCP8.5). We selected 2070 as the targeting year to compare with the current climate condition. Table 1 summarizes the storm surge and RSLR scenarios investigated in this study. Note that the current climate condition refers to water level and climate conditions in year 2020.

Table 1. Relative sea level rise and storm surge scenarios (NAVD88) prepared for the storm surge model (Tahvildari and Castrucci, 2020). Water levels are at the Sewells Point tide gauge in Norfolk, VA.

| Year | Storm Surge Return Period (yr) | RSLR Scenario |
| --- | --- | --- |
| 2020 | 10 (1.55m), 50 (2.07m) | No RSLR |
| 2070 | 10 (1.55m), 50 (2.07m) | High (1.471m) |

2.6.2 Rainfall under Climate Change Scenarios

The impact from the change of rainfall intensity due to climate change is explored in this study. Changes in rainfall intensity are one of the primary impacts of climate change in urban areas (Bi et al., 2015). Changes in other factors, for example, temperature and wind speed, are not simulated in this study. Following the example of Morsy et al., (2019), the changes in rainfall intensity in Virginia are estimated under multiple climate change scenarios. Generally, the future intensity-duration-frequency (IDF) curves for the study area are estimated by analyzing future rainfall projections. Then, rainfall scenarios are generated based on these IDF curves and assumed rainfall distribution.

Rainfall projections from global climate models (GCMs) or regional climate models (RCMs) are commonly used to estimate the impacts of climate change. The spatial resolutions of GCMs are generally coarse, i.e., greater than 100km (Bi et al., 2015). The spatial resolutions of RCMs are typically in between 12km to 50km (Arnbjerg-Nielsen et al., 2013; Bi et al., 2015). In this study, the rainfall projections were collected from the Coordinated Regional Climate Downscaling Experiment (CORDEX, 2019), which is a global repository of RCM simulations covering all continents of the globe. In CORDEX, climate projections are downscaled using dynamic RCMs for 14 domains with the IPCC 5th GCMs outputs as the boundary condition. The CORDEX products in North America are available for the Representation Concentration Pathway (RCP) 4.5 and RCP 8.5 climate scenarios. The spatial resolutions of these products vary from 22km to 44km. In the current study, the daily rainfall projections for the period of 1950 to 2100 were extracted at the Norfolk International Airport. Then, the rainfall projections were bias corrected using historical observations from 1950 to 2005 by using the modified Empirical Quantile Mapping (EQM) method proposed by Themeßl et al. (2011). In the EQM method, the historical observation and hindcast simulation from 1950 to 2005 were used to build a quantile-quantile mappings (QQMs). Next, the QQMs were applied to correct the bias in the precipitation projections.

The bias corrected rainfall projections were then used to define the IDF curves. Because future rainfall is non-stationary under a changing climate, IDF curves are not fixed. However, an assumption in this study is that the IDF curves are fixed for the periods of interest. The same statistical methods and probability distribution, Generalized Extreme Value (GEV), used to construct the NOAA Atlas 14 IDF curves (Bonnin et al., 2006) were chosen to construct the IDF curves used in this study. Shen et al. (2019) summarized and analyzed the rainfall durations of large storm events observed at the Norfolk International Airport. This study found that majority of these large storms have rainfall durations of about 24 hours. Thus, the current study focuses on the 24-hour rainfall event to represent large storm events. The rainfall intensities for a 24-hour duration storm event with return periods varying from 1 to 100 years were estimated under RCP4.5 and RCP8.5 scenarios for the period 2056-2085 (simply stated as the 2070 IDFs in the remainder of this paper). Figure 5 shows the comparison between the 2070 IDFs and the current IDF obtained from NOAA Atlas 14 (https://hdsc.nws.noaa.gov/hdsc/pfds/pfds_map_cont.html).

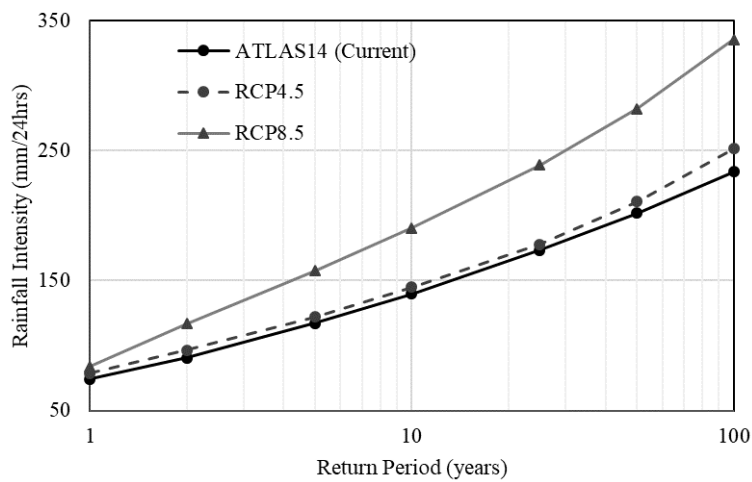

Figure 5. Rainfall intensities for a 24-hour duration storm from Atlas 14, representing the current period, and for 2070 estimated under both the RCP 4.5 and RCP 8.5 scenarios.

According to Merkel et al. (2015), the study domain is located in the region of NOAA Type-C rainfall distribution zone. Therefore, we adopted the NOAA Type-C rainfall distribution for creating synthetic rainfall hyetographs for the 24-hour rainfall scenarios. We used the 10-year and 50-year rainfall scenarios to model flooding because they represent significant, but not catastrophic, flood impacts to the study area. We choose the RCP8.5 scenarios to demonstrate the potential impact of these more significant climate change scenarios on flooding.

To explore the combined impact of rainfall and storm tide under both the current and future climate conditions, the designed storm tide and rainfall scenarios were put into a series of inputs to the coupled flood model. Shen et al. (2019) investigated the time lag between storm surge and rainfall and showed how it would significantly influence the impact of flooding in the study region. Different time lags were not explored in this study. Instead, the time lag was set to be the same as what was observed for Hurricane Irene (2011) where the rainfall peak was 8 hours ahead of tide level peak. An example model inputs for a 10-year rainfall and a 10-year storm surge event in 2020 is provided in the Figure A1 of Appendix A.

## 2.7 Assessing Transportation Impacts

The flood simulations outputs are used to estimate impacts to the transportation network within the study area. In this study, the impact on roads is described as road closures and the criterion for closing a road is assumed to be a flood depth of 0.3m or more on the roadway, as predicted by the flood model. This flood depth is a conservative road closure criteria because it is the height of air inlets of most vehicles (Yin et al., 2016)

The transportation network was obtained from the City of Norfolk (https://www.norfolk.gov/1596/Geographic-Information-Systems) and the traffic data of major roads was gathered from Virginia Department of Transportation (VDOT: https://www.virginiadot.org/info/ct-TrafficCounts.asp). The study domain contains 6983 road links and 450 major road links, as shown in Figure 6. The annual average daily traffic (AADT) of the major roads varies from 60 to 151,000. About 60% of the major road links AADTs are greater than 10,000.

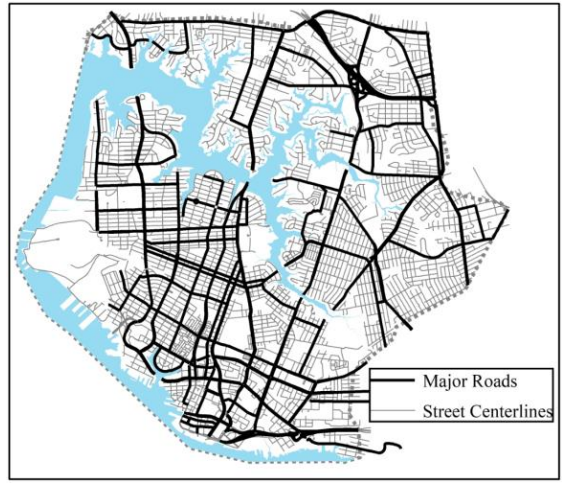

Figure 6. Transportation network inside the study domain.

## 3. Results and Discussion

*3.1 Model Evaluation*

The storm surge model was built to assess the impact of sea level rise in Norfolk and validated by observations from two tide gauges during Hurricane Irene (2011) in Castrucci and Tahvildari (2018) and Tahvildari and Castrucci (2020). For detailed information about the storm surge model validation, please refer to these studies. Here we focus on evaluating the performance of the coupled flood model, rather than just the storm surge or urban flood model in isolation.

A USGS gauge (latitude: 36.8588, Longitude: -76.2986) was temporarily deployed in the Hague Community for the Hurricane Irene (2011) storm event at the location shown in Figure 7a. A comparison between the water level observations at this USGS gauge and the output from the coupled model is provided in Figure 7b. Overall, the simulation shows a close match with the observation in both phase position and magnitude. The coupled model slightly underestimates the peak water level (by 0.02m) and has a positive phase shift (of about 30 minutes) compared to observations. Given that this study focuses more on the peak flood impact than the timing of these impacts, these results were deemed acceptable based on this evaluation.

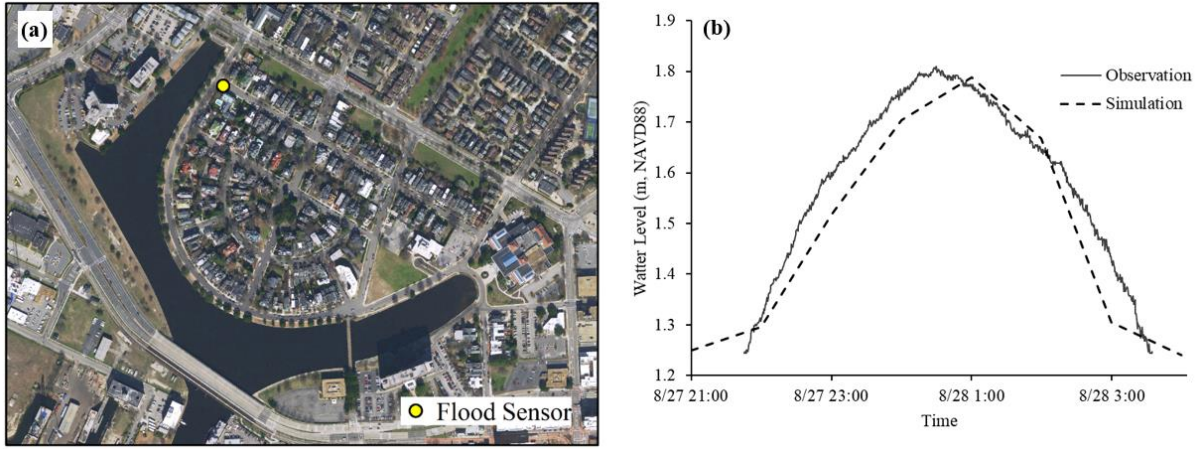

Figure 7. Evaluating the coupled flood model with a temporary USGS gauge deployed during Hurricane Irene (2011). (a) Location of the USGS gauge; (b) Comparison between observed and simulated water levels.

Given that 2D flood models provide flood inundation map outputs, we also evaluated the flood model using areal imagery data. As shown in Figure 8, flood extents were extracted from drone footage at three locations inside the study domain for the Hurricane Matthew flood event, as described in the methods section. Area 1 is directly connected to the tidally influenced Lafayette River, and flooding is therefore controlled by the downstream tidal boundary condition. Area 2 is about 700m away from the shoreline and is under the combined impact of tidal flooding and pluvial flooding, as shown in Shen et al. (2019). Area 3 is located at the shoreline and exposed to direct tidal flooding. Therefore, even though flood extents are only available at three local regions, they represent areas where tidal, pluvial, or tidal and pluvial flooding are at play and can serve as valuable evaluations of the overall model system.

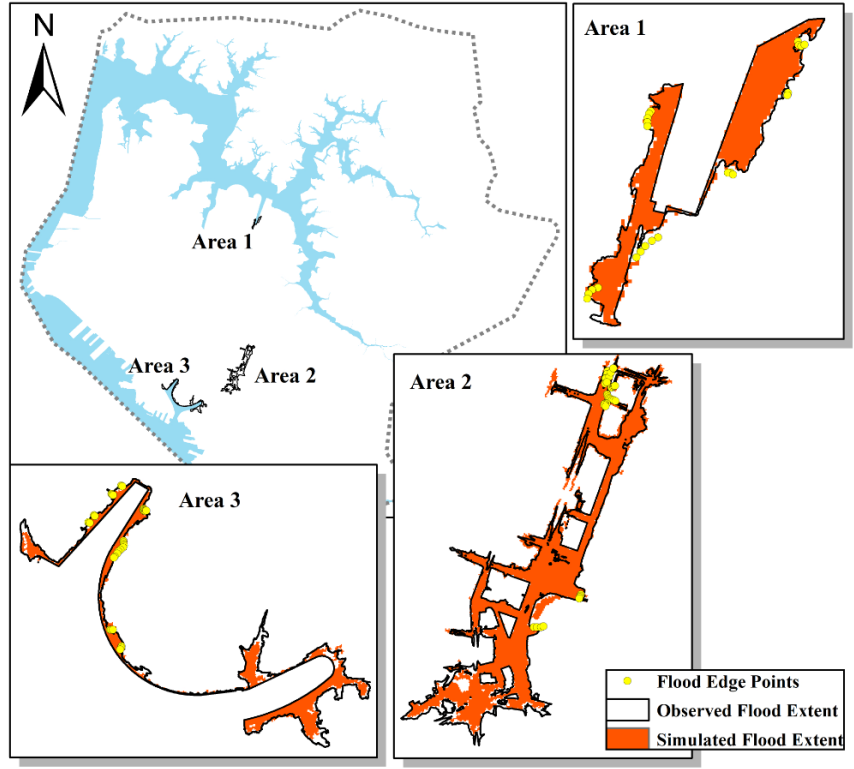

Figure 8. Flood extents extracted from drone footages.

Precision and recall were adopted as accuracy metrics to evaluate the performance of the coupled flood model. Precision and recall are commonly used metrics to compare the relevance between model predictions and ground truth for image classification. Recall represents the proportion of flood extent that was correctly identified by the model compared to the ground truth flood extent, taken from the imagery, and is defined as

$$Recall = \frac{True\ flood\ extent\ simulation}{Total\ true\ flood\ extent} \qquad (3)$$

Precision represents the proportion of flood extent that was correctly identified by the model compared to the total flood extent produced by the model and is defined as

$$Precision = \frac{True\ flood\ extent\ simulation}{Total\ flood\ extent\ simulation} \qquad (4)$$

The flood extent precision and recall in each evaluation area are shown in Table 2. Figure 8 and Table 2, show that the spatial distribution of the simulated flood extents match well with the flood extents extracted from the images. The overall precision across these three areas varies from 0.71 to 0.92 and the recalls varies from 0.66 to 0.84. Within these three areas, the model shows the best performance in Area 1 with precision and recall of 0.92 and 0.84, respectively. In Area 2, the topography and mechanisms of flooding are more complex than the other two areas. Figure 8 illustrates that most of the underestimated flood extent in Area 2 is in the southwest of that area that includes the parking lot of the Virginia Opera and intersections of three major roads. The model shows the lowest precision and recall in Area 3, which faces direct tidal flooding. The underestimation is most pronounced in the eastern portion of Area 3. There is no direct observation or footage available for this portion of the study region, so the flood extent had to be estimated from GIS analysis using digital terrain data. Further investigation is needed to better understand flooding in this area and if the model is underestimating flooding in this region. Despite these challenges and opportunities for improvement, the model shows good performance overall across the three evaluation areas as judged by precision and recall values.

Table 2. Flood extent precision and recall in the three evaluation areas.

|  | Precision | Recall |
|---|---|---|
| Area 1 | 0.92 | 0.84 |
| Area 2 | 0.88 | 0.74 |
| Area 3 | 0.71 | 0.66 |

*3.2 Comparison with the Bathtub Method*

This section presents results of the comparison between the coupled flood model and the bathtub method for assessing the impact of SLR and storm tide in coastal regions. Results from the 10-year storm tide events in current (2020) and future (2070) sea level conditions are taken as a demonstration (Figure 9). Similar to results shown by Castrucci and Tahvildari (2018), we found that the bathtub method overestimates the flooding extent for regions topographically and hydraulically connected to open water. In Figure 9, the bathtub method overestimated flood extent near the shoreline by 9.5% and 3.1% in 2020 and 2070, repsectively. However, due to the existence of underground drainage systems in the coupled flood model, several inland local depression areas are simulated by the model to be flooded by backward pipe flow from high tide level but not by the bathtub method. Compared to the bathtub method, the flood model estimated 9.0% and 4.0% more flood extent in the inland region in 2020 and 2070, respectively. This is because the bathtub method does not account for rainfall-driven flooding, which in combination with the backwater pipe flow, is responsible for this inland flooding.

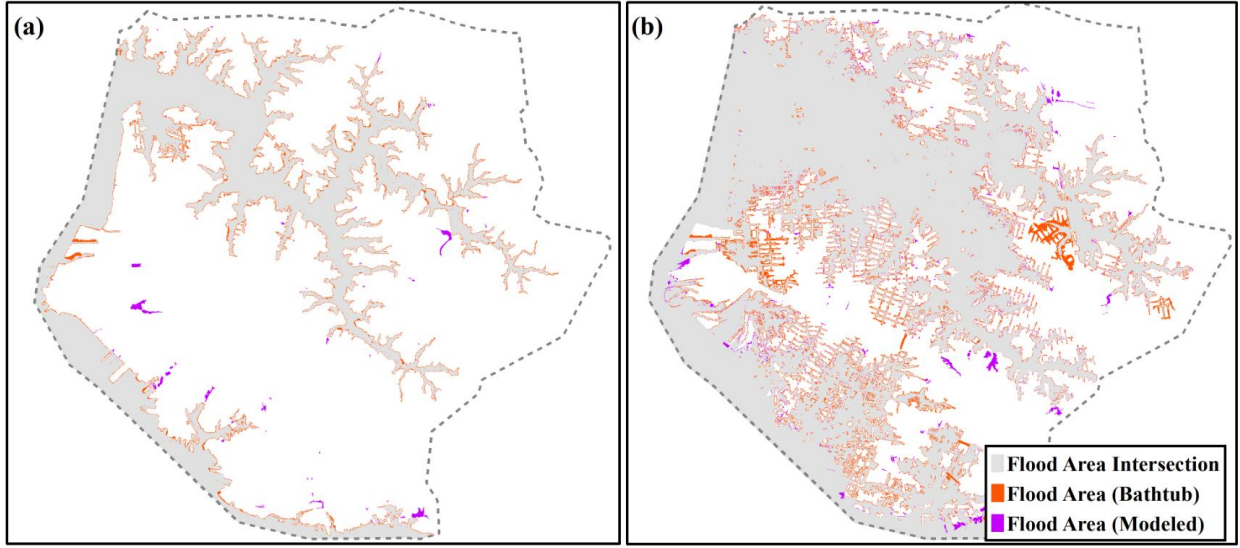

Figure 9. Demonstration of flood area estimation using the bathtub method and the coupled modeling system in two scenarios: (a) 10-year storm surge under the 2020 climate conditions and (b) 10-year storm surge under the 2070 high RSLR scenario.

The coupled flood model and bathtub method are also compared for other storm tide scenarios under the 2020 and 2070 sea level conditions. For this comparison, we used the percent difference in flood area as follows.

$$Diff = \frac{Flood\ Area_{bathtub} - Flood\ Area_{modeled}}{Flood\ Area_{bathtub}} \times 100\% \qquad (5)$$

From these results (Figure 10), we can draw the following insights. First, the overestimation of the bathtub method increases as the storm return period increases. Second, when the tide peak level is lower than a certain level, for example in the "no storm surge" 2070 scenario in Figure 10, the total flood area from the coupled flood model can be greater than the bathtub method. In this scenario, the total inland area flooded by backward pipe flow exceeds the overestimation of the bathtub method in the near coastline regions.

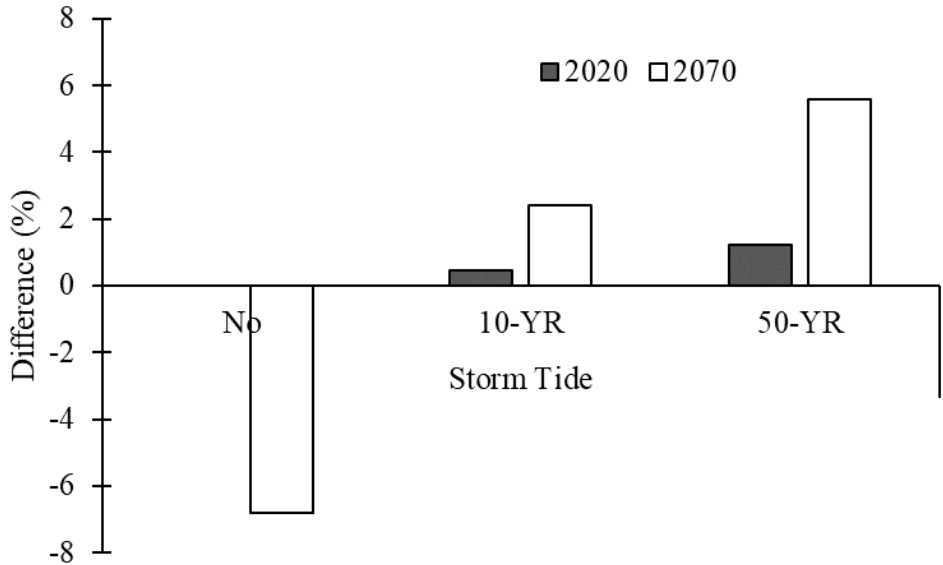

Figure 10. Comparison between flood area estimates from the bathtub method and the coupled flood model under the 10-year and 50-year storm surge events in 2020 and 2070 (no rain).

*3.3 Flood Areas*

The flood maps for all compound storm events under both the 2020 and 2070 climate conditions are presented in Figures 11 and 12. As expected, tidal flooding primarily occurs near coastline and tidal rivers with low-lying topography. In Figure 11, under the no rainfall condition, several inland areas are flooded by the 10-year and 50-year storm tide. Flooding in these local depression areas seem to be caused by backward pipe flow because these stormwater pipelines in this region lack a tide gate or flap gate to prevent backward flow. Figure 11 shows how pluvial flooding is distributed across the inland region, specifically gathering in areas that lack effective drainage infrastructure. In the analyzed rainfall scenarios from Figure 12, the flood extents are estimated to increase dramatically due to the increase of rainfall intensity and SLR in 2070. Under the no storm surge condition in 2070, a large portion of the coastline region is estimated to be flooded due to the increase of the base tide level alone. When impacted by storm tide on top of SLR, majority portion of the study domain is expected to be flooded.

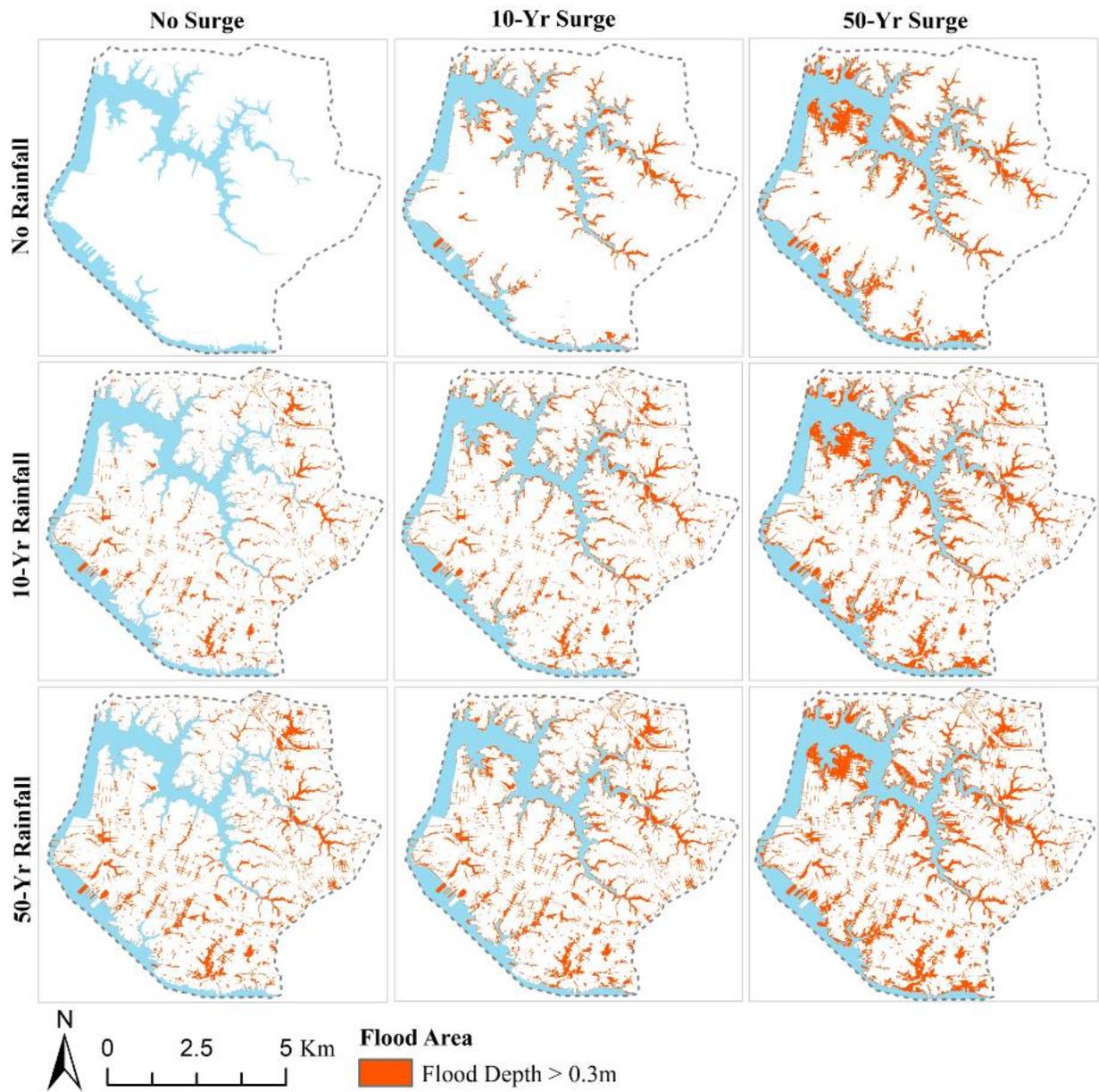

Figure 11. Flood maps for different compound storm surge scenarios under current climate and sea level conditions.

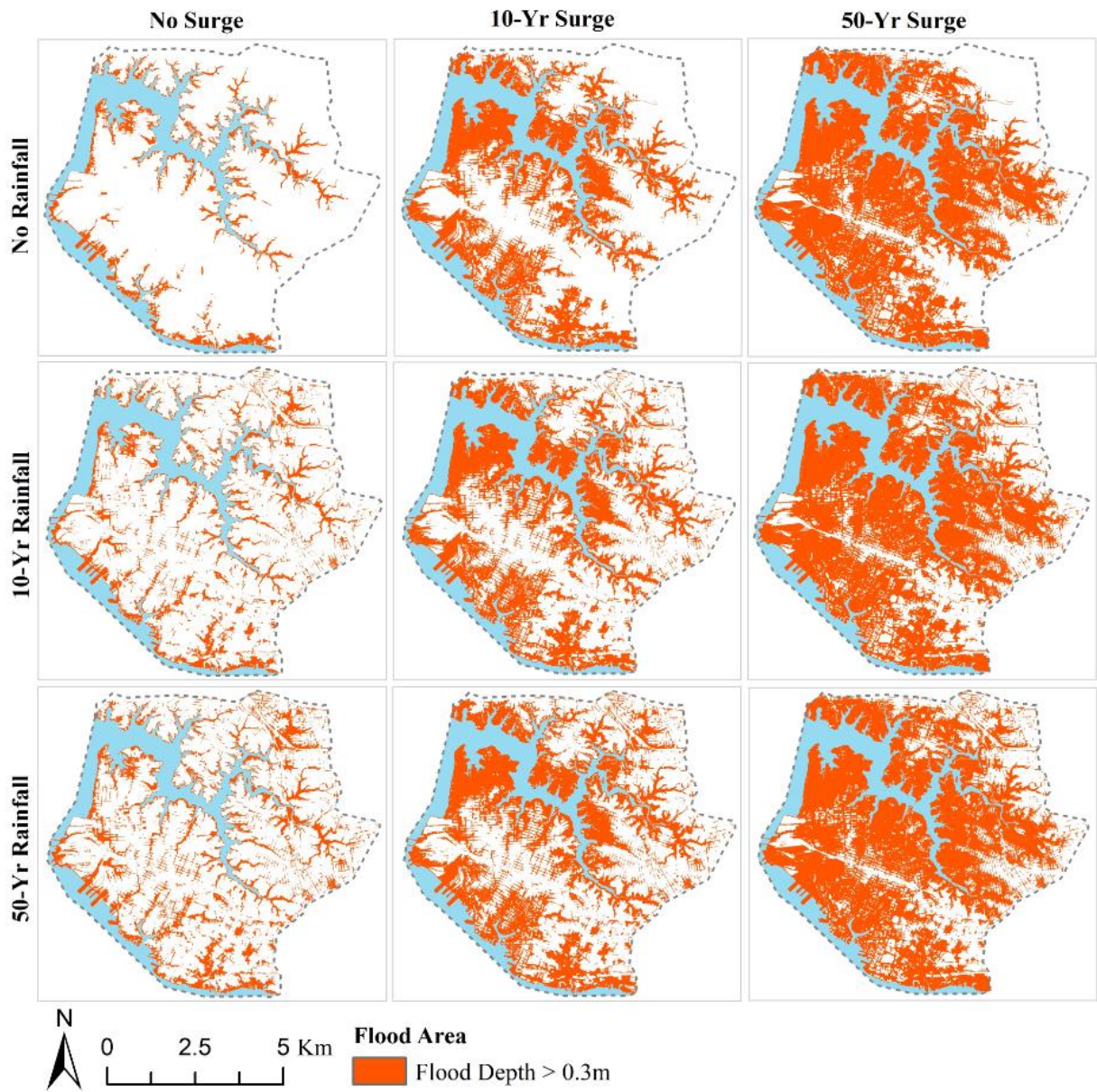

Figure 12. Flood maps for different compound storm surge scenarios under 2070 climate and sea level conditions

Figure 13 shows the percentage of the total flooded area for all analyzed compound storm scenarios. Under the 2020 condition, the compound storm events are estimated to flood 7% to 20% of the land area, depending on the storms return period. Under the projected 2070 conditions, the compound storms would flood 18% to 66% of the land area under these same return period storms. From Figure 13b, a so called "sunny-day flood" scenario (just high tide with no rainfall or storm tide) in 2070 is still estimated to flood 11% of the land area. This flood extent is nearly equivalent to a 50-year (13%) storm tide in 2020. From 2020 to 2070, the flood extent increase by about 11% on average for both the 10-year and 50-year rainfall scenarios under the no storm tide condition. That increase is primarily caused by the combined effect of rainfall intensity increases and RSLR causing backwater pipeline flooding and reduced pipeline capacity. Over the same time span, even with no rainfall and only storm tide, the flooded extent increase by 32% and 49% for the 10-year and 50-year storm tides compared to the no storm surge scenario, respectively. Taking the 50-year storm tide in 2070 as an example, storm tide alone will flood 62% of the land area. Combined with a 50-year rainfall event, the compound storm is estimated to flood 66% of the land area, which means only 4% of the total flooding in this scenario is attributed to rainfall. Therefore, tidal flooding has a greater impact on flooding compared to pluvial flooding in the study domain. under project RSLR scenarios.

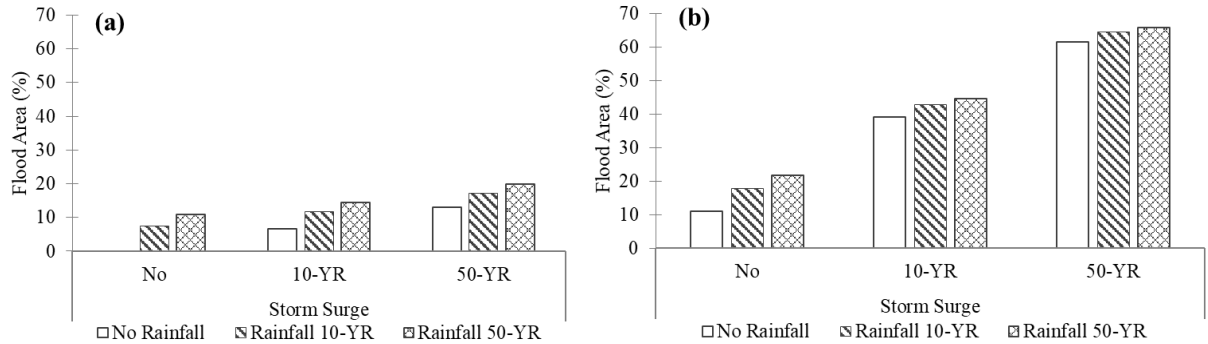

Figure 13. Percentage of total flood extent (flood depth > 0.3m) for (a) 2020 and (b) 2070 under a high SLR scenario and RCP8.5 climate change scenario.

*3.4 Flood Impact on the Transportation Network*

The coupled flood model provides a method for estimating flood impact on critical urban infrastructure systems. As a demonstration, the flood model is applied to assess the flood impact to transportation networks under a set of storm scenarios.

The impact to the transportation network is quantified as a percentage of flooded road length to total road length in the study region (Figure 14). Under 2020 conditions, the analyzed storm events would flood between 4.6% (10-year storm surge alone) to 22.3% (50-year storm surge and 50-year rainfall) of roadways. In 2070, the flooded roadways are estimated to vary from 19.0% (10-year rainfall alone) to 66.9% (50-year storm surge and 50-year rainfall). Meanwhile, on a sunny day scenario (high tide alone with no rainfall or storm surge) in 2070, 5.2% of the road length would be flooded. This sunny day or nuisance flooding projected for 2070 is similar to the impacts of a 10-year storm tide in 2020.

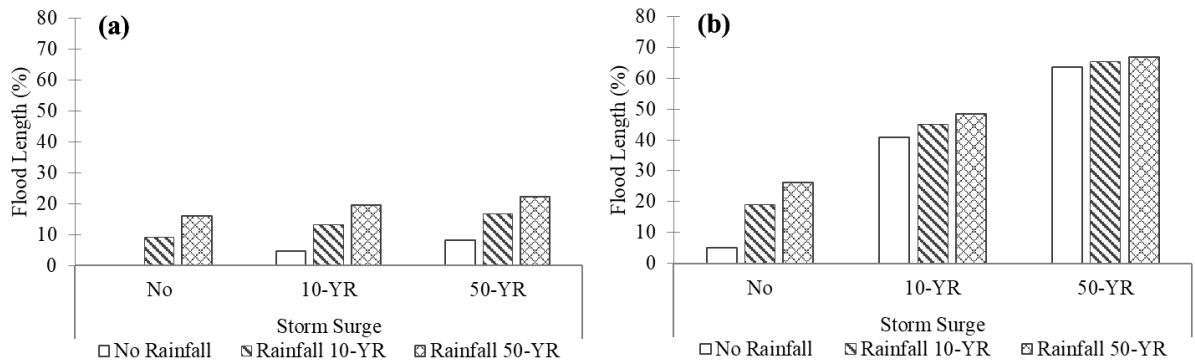

Figure 14. Percentage of flooded road length under the tested storm scenarios: (a) 2020 conditions and (b) 2070 conditions under high SLR and RCP8.5 scenarios.

One major advantage of a dynamic flood model, compared to static methods like the Bathtub method, is its ability to simulate the duration of flooding rather than just the peak flooding impacts. Figure 15 shows the percentage of roadway links that are flooding during storm scenarios. The rainfall peak is 8 hours ahead of the tide level peak in these scenarios, as explained in the methods section, resulting in two peaks for the roadway flooding impacts. Under 2020 conditions (Figure 15a), storms with both significant rainfall and storm tide result in similar impacts during the start (rainfall dominate) and ending (storm tide dominate) portions of the storm event. However, under 2070 conditions (Figure 15b), we see again that storm tide because a more impactful flooding mechanism. In this scenario, a storm with both significant rainfall and storm tide would suffer the most impact during the portion of the storm where storm tide was the dominate flooding mechanism. Looking at the impact from storm scenarios with only rainfall-driven flooding, the maximum number of flooded road links is projected to increase by 8.7% and 14.8% under the impact of a 10-year and a 50-year rainfall events, respectively, from 2020 to 2070. For a tidal-driven flooding scenario with no rainfall, the maximum percentage of flooded road links will increase by 51.7% and 62.4% for a 10-year and a 50-year storm tide events, respectively. Thus, this suggests the impact of flooding on the transportation network will be most impacted by tidal-driven flooding compared to rainfall-driven flooding under the projected RLSR and climate change scenarios.

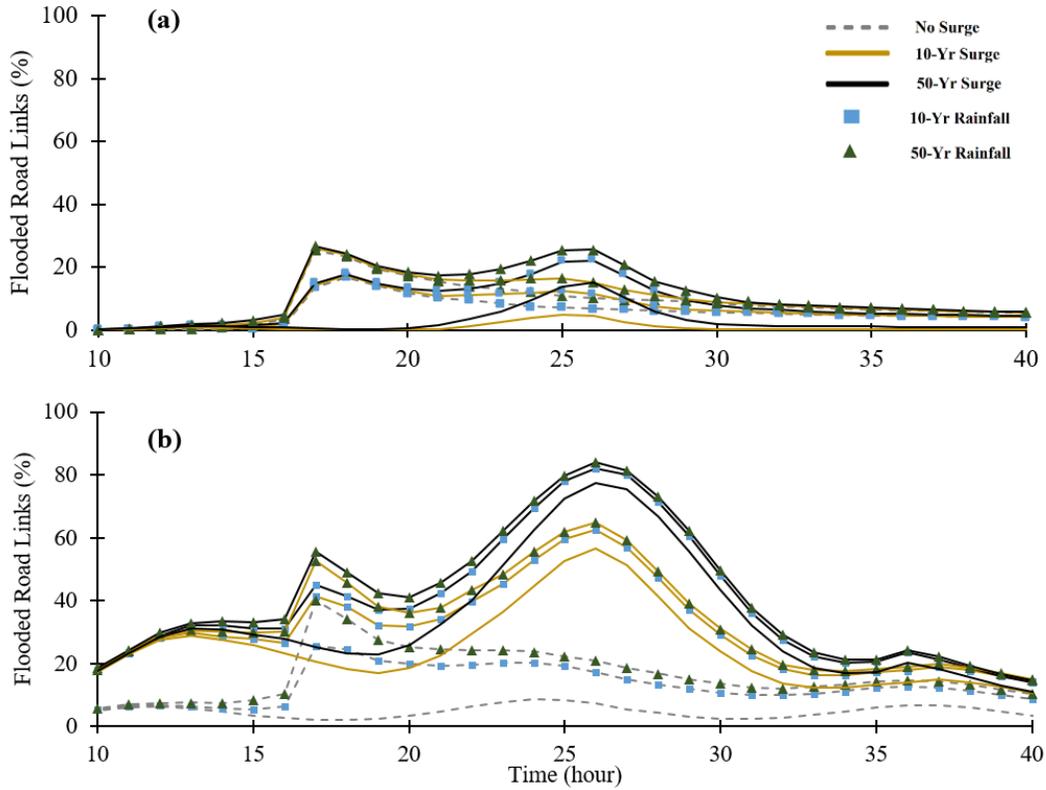

Figure 15. Percentage of flooded road links in time series during the simulation periods of all the tested compound storm scenarios: (a) current climate conditions and (b) 2070 high SLR scenario and RCP8.5 climate change scenario.

To summarize the spatial-temporal impact on the transportation network, a metric named the total link closed time (TLCT), is introduced. TLCT is defined as

$$TLCT = \frac{\sum_i^N T_{closed,i}}{T_{total} \times N} \times 100\% \tag{6}$$

where, N is the total number of road links, $T_{closed,i}$ is the total closed (due to flooding) time of link $i$ during a storm event, and $T_{total}$ is the total length of the simulation period. TLCT for the road links in the study domain was computed for all storm scenarios, and the results are shown in Figure 16. TLCT is sensitive to the increase of both rainfall intensity and storm surge under the 2020 conditions, as illustrated in Figure 16a. Under these conditions, a 10-year rainfall would result in TLCT of 4.2%, but a 50-year storm surge would only cause TLCT of 1.8%. This is because the duration of pluvial flooding is longer than tidal flooding in the study region under current conditions. However, these conditions are expected to change by 2070. As shown in Figure 16b, the 10-year and 50-year storm tide events will result in TLCTs of 16.9% and 23.6%, respectively. Interestingly, adding the impact of rainfall to these scenarios only increases the TLCT by 1.3% to 4.1%. Therefore, storm tide events are expected to be the dominant factor interrupting the transportation network in 2070. Finally, sunny day or nuisance flooding is expected to cause a TLCT of 4.6%, which is more than two times that of a 50-year storm surge in 2020. Considering the duration of flooding and its impact on transportation systems, nuisance flooding in 2070 is likely to be a serious challenge for transportation management unless efforts are made to prevent tidal flooding impacts, such as preventing backflow of tidal water through subsurface drainage networks.

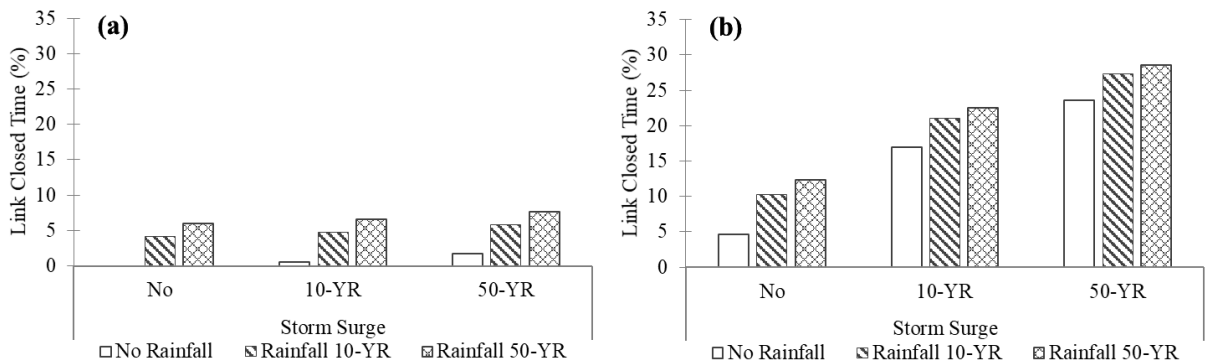

Figure 16. Total links closed time (TLCT) for all road links in the study domain under the analyzed storm scenarios: (a) 2020 conditions and (b) 2070 conditions with high SLR and RCP8.5 climate change scenarios.

## 4. Conclusions

The overarching objective of this study is to advance methods for assessing the combined impact of rainfall and storm tide on coastal cities under a changing climate. As a case study and example application of the methods, a coupled flood model consisting of a 1D stormwater pipe/2D overland flood model and a 2D storm surge model was built for a large portion of Norfolk, VA. The coupled flood model can simulate the hydrodynamic and wave processes of a storm tide in the ocean and its progression on land, as well as well as pluvial flooding due to excess rainfall and resulting runoff. By simulating the complex stormwater drainage infrastructure system as well in the model, the coupled flood model is especially suited to analyzing flooding in an urban environment. This detailed numerical model can generate simulations with high spatial-temporal resolution to support flood hazard assessment and management. We made use of available point observations and imagery data to evaluate the model performance, and created an approach to extract flood extent from drone imagery to evaluate the model. The model was compared to the more commonly used bathtub method for flood hazard assessment and the utility of the model is demonstrated by quantifying flooding impacts to the transportation network within the region.

Results of the comparison of the detailed flood model with the simpler bathtub method show that the bathtub method overestimated the flooding extent near shorelines by 9.5% and 3.1% for a 10-year storm tide event in 2020 and 2070, respectively. It, however, underestimated flooding extent for inland areas by 9.0% and 4.0%, respectively, for the same events. As described in prior work, the error in the bathtub method can be attributed to neglecting several important physical processes of coastal flooding: (1) the effect of landscape roughness; (2) short-term dynamics of flows; (3) conservation of mass for flows; (4) existing drainage infrastructure (Ramirez et al., 2016; Tahvildari and Castrucci, 2020). The bathtub method uses high water levels at tidal gauges, which are generally located in deeper water and their measurements can differ in both amplitude and phase from water level at shorelines or overland. Additionally, the bathtub method assumes that maximum tide levels are maintained for an indefinite duration. As a result, this study adds to the literature suggesting that the bathtub method consistently overestimates flood extent near the shoreline (Ramirez et al., 2016; Tahvildari and Castrucci, 2020). This work further contributes to existing literature showing how urban stormwater drainage networks function within coastal flooding and that simpler hazard assessments, like the bathtub method, can underestimate flooding within inland areas by not including this infrastructure. Seawater can backflow through drainage pipes and daylight in inland areas. Furthermore, rainfall-driven flooding can flood inland areas and, due to drainage infrastructure backflow, cause increased flooding.

This work also demonstrates the value of time-dependent flood information possible from a dynamic coupled flood model as opposed to static, time-independent approaches. By capturing the temporal dynamics of flooding, this work shows how pluvial flooding can result in a longer flood duration than tidal flooding for storms with the same recurrence interval due to the time required to drain rainfall from the system. Therefore, pluvial flooding can cause a larger interruption to the transportation network than tidal flooding under current conditions when the time duration for flooding is taken into account. This illustrates how dynamic flood models can be insightful for understanding flood hazards to critical urban infrastructure systems. As another example, we used a metric called total link close time (TLCT) to assess the impact to the transportation network within the study region. Based on this analysis we found, based on model projections, that sunny day or nuisance flooding in 2070 will cause a 4.6% TLCT. For context, this is more than two times the impact a 50-year storm tide would cause under current conditions (1.8% TLCT). Comparing the magnitude of flood extent and duration, we found that while rainfall and tidal flooding have similar impacts today, by 2070 storm tide will be the more dominant mechanism for causing flood impacts due to SLR impacts outpacing increased rainfall impacts. By 2070, the model projects that a 50-year storm surge event on top of the high SLR scenario is projected to flood 62% of the study domain.

The coupled flood model used in this study was built with many assumptions that could be tested and advanced in future work. These assumptions are mentioned in the methods section, but a few are highlighted here as those we consider to be the highest priority. First, the model assumed saturated soil conditions due to the large portion of impervious surfaces, high groundwater table, and focus on more significant storm events (e.g., 10-year and great return periods). Implementing the infiltration and groundwater components into the coupled model could help to test this assumption and would make the model more applicable to less intense, more frequently reoccurring storm events. Second, the model has a relatively simple technique for coupling the ocean and overland models where, essentially, the ocean model provides a tide water level boundary condition for the overland model. In future work, tide and velocity could be transferred between these two models to improve the physics of the coupling or, more ambitiously, a full two-way coupling between these models could be explored. The model runtimes and spatiotemporal mismatch between these two models, however, will make a full coupling extremely challenging. Lastly, the model assessment made use of limited data including a tide gauge and imagery data collected for one storm event using a drone. While this is a fairly standard approach for model evaluation, ideally much more data including both point observations and imagery would be available and used to more fully evaluate and calibrate the model across a variety of storm events.

**Acknowledgements**

This work was supported by the National Science Foundation under the award number 1735587. The authors wish to acknowledge the BMT for the TUFLOW HPC license and kindly help on model building and problem solving. We also would like to thank for Marlene McGraw's help and advice on editing the manuscript.

**Appendix A.**

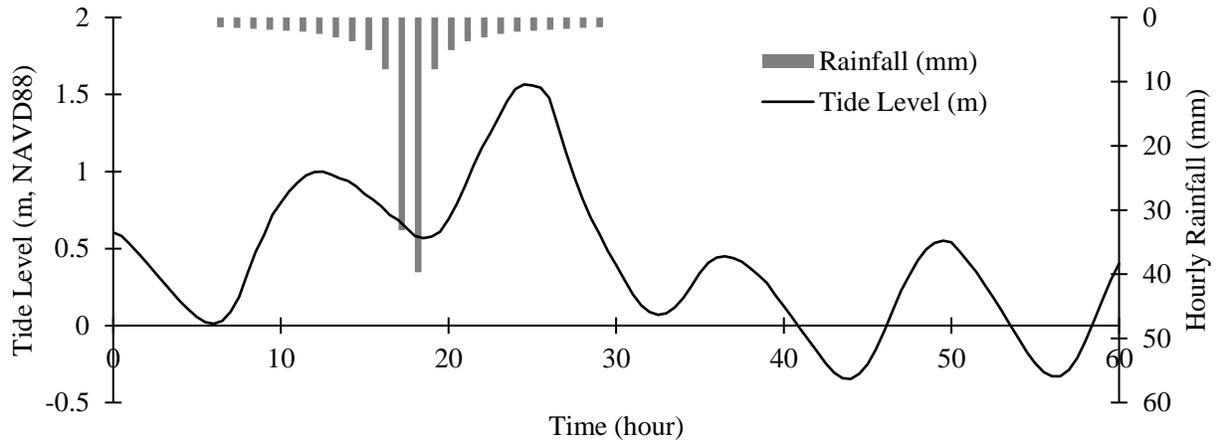

Figure A1. Designed compound storm scenario consists of a 10-year rainfall and a 10-year storm surge events. Note that the tide level is at the Sewells Point tide gage in Norfolk, VA.